\documentstyle[12pt,aaspp4,epsf]{article}     

\slugcomment{Accepted for publication in ApJL}

\begin{document}

\title{
The X-ray Luminosity Function of Bright Clusters in the Local Universe
}  
  
\vskip 30pt
\author{S. De Grandi\altaffilmark{1,2}, 
        L. Guzzo\altaffilmark{2},
        H. B\"ohringer\altaffilmark{1},
        S. Molendi\altaffilmark{3},
        G. Chincarini\altaffilmark{2,4},
        C. Collins\altaffilmark{5},
        R. Cruddace\altaffilmark{6},
        D. Neumann\altaffilmark{1,7},
        S. Schindler\altaffilmark{1,5},
        P. Schuecker\altaffilmark{1}, and
        W. Voges\altaffilmark{1}
}

\altaffiltext{1}{Max-Planck-Institut f{\"u}r extraterrestrische Physik,
Giessenbachstra$\ss$e 1, 85740 Garching bei M\"unchen, Germany}
\altaffiltext{2}{Osservatorio Astronomico di Brera, via Bianchi 46,
22055 Merate (LC), Italy}
\altaffiltext{3}{Istituto di Fisica Cosmica, Consiglio Nazionale
delle Ricerche, via Bassini 15, 20133 Milano, Italy}
\altaffiltext{4}{Universit\`a di Milano, via Celoria 16,
20133 Milano, Italy}
\altaffiltext{5}{Astrophysics Research Institute, Liverpool John-Moores
University, Byrom-Street, Liverpool L3 5AF, United Kingdom}
\altaffiltext{6}{E. O. Hulburt Center for Space Research, Naval Research
Laboratory, Code 7620, 4555 Overlook Ave., Washington, DC 29375}
\altaffiltext{7}{Commissariat a l'Energie Atomique/Saclay, 
Service d'Astrophysique, L'Orme des Merisiers 
Bat.709, 91191 Gif-sur-Yvette, France}

\begin{abstract}
We present the X-ray luminosity function (XLF) for clusters of galaxies
derived from the RASS1 Bright Sample.
The sample, selected from the $ROSAT$ All-Sky Survey 
in a region of 2.5 sr within the southern Galactic cap, 
contains 130 clusters with flux limits   
in the range $\sim 3-4\times 10^{-12}$ ergs cm$^{-2}$ s$^{-1}$ in the 
0.5-2.0 keV band.  
A maximum-likelihood fit with a Schechter function of the XLF 
over the entire range of luminosities ($0.045 - 28.\times 10^{44}$ 
ergs s$^{-1}$), gives 
$\alpha = 1.52^{+0.11}_{-0.11}$, 
$L_* = 3.80^{+0.70}_{-0.55}\times 10^{44}$ ergs s$^{-1}$, and 
$A = 5.07\pm 0.45\times 10^{-7}$ Mpc$^{-3}$ ($10^{44}$ 
ergs s$^{-1}$)$^{\alpha-1}$.
We investigate possible evolutionary effects within the sample, out 
to our redshift limit ($z\sim 0.3$), finding no evidence for evolution.  
Our results are in good agreement with other local estimates of the XLF,
implying that this statistic for the local universe is now  well determined. 
Comparison with XLFs for distant clusters ($0.3 < z < 0.6$),
shows that no evolution is present for $L_X \lesssim 10^{44}$ 
ergs s$^{-1}$.
However, we detect differences at the $3\sigma$ level, between 
our local XLF and the distant one estimated by Henry et al. for the 
EMSS sample. 
This difference is still present when considering the EMSS sample revised 
by Nichol et al.

\end{abstract}

\section{Introduction}
Clusters of galaxies have been extensively investigated as a powerful
tool for cosmological studies.
The X-ray luminosity function (XLF) is one of the most studied properties
because it is related to the cluster mass function, 
gives information on the amplitude of the cosmic density  fluctuation power 
spectrum and is sensitive to cluster evolution.
Deep surveys covering small solid angles give information essentially on
the faint-end of the XLF and on its redshift dependence 
(e.g., Henry et al. 1992, Rosati et al. 1998, Collins et al. 1997,
Romer 1998, Jones et al. 1998).
In contrast, shallow, wide-angle samples, allow the determination of the 
``local'' (i.e., $z\lesssim 0.3$) XLF over the entire cluster luminosity 
range, 
which is crucial for studies of cluster evolution (e.g.,  
Briel \& Henry 1993, Ebeling et al. 1996, Ebeling et al. 1997).
Early XLF studies of flux-limited samples compiled from $EXOSAT$ and 
$Einstein$ data (Edge et al. 1990, Gioia et al. 1990, Henry et al. 1992)
showed evidence of negative cluster evolution 
at $z\gtrsim 0.1-0.2$ (Edge et al. 1990) or $z\gtrsim 0.3$ (Henry et al. 
1992),
whereas more recent work indicates that no evolution is present, at least 
for $z\lesssim 0.8$ and X-ray luminosities lower than $\sim 3\times 
10^{44}$ ergs s$^{-1}$ (e.g., Burke et al. 1997, Rosati et al. 1998,
Vikhlinin et al. 1998a).

To exploit the unique opportunity provided by the $ROSAT$ All-Sky Survey 
(RASS), we have constructed, from the first processing of the survey data,  
the RASS1 Bright Sample (De Grandi et al. 1999, hereafter Paper II) 
of clusters of galaxies, which covers a contiguous  area of 8235 
deg$^2$ in the southern hemisphere ($b_{II}<-20^o, \delta<2^o.5$).
This sample was constructed as part of an ESO Key Programme 
(Guzzo et al. 1995) aimed at surveying 
all the southern-sky RASS cluster candidates.  This is now known as
the REFLEX cluster survey (B\"ohringer et al. 1998), and it is currently 
nearing completion.
Our sample contains 130 clusters  with $ROSAT$ PSPC hard-band ($0.5-2.0$ keV) 
count rates higher than 0.25 counts s$^{-1}$ 
(corresponding to flux limits ranging 
from 3.05 to $4\times 10^{-12}$ ergs cm$^{-2}$ s$^{-1}$).
A comprehensive discussion of the analysis of the X-ray data is given in 
De Grandi et al.(1997, hereafter Paper I), whereas the sample selection 
function and the estimation of the overall completeness and biases
are presented in detail in Paper II. 
The relatively high flux limits (see Paper II), the large sky area 
covered, and the redshift distribution of the RASS1 Bright Sample make 
this sample particularly useful to study the local XLF out to $z\sim 0.3$.
A first estimation of the XLF from a preliminary version of this sample, 
was presented by De Grandi (1996). 
In this Letter, we compute the XLF from the definitive sample described in 
Paper II.
Presently, the only other sample selected from the RASS1 data
with characteristics similar to those of our sample is the brightest 
cluster sample (BCS) of Ebeling et al. (1997, 1998).
However, the selection procedures applied to compile the two  
samples are completely different with respect to both the selection procedure 
for the cluster candidates and the technique used to analyze the X-ray data.
Throughout the Letter, we assume $H_0 = 50$ km s$^{-1}$ Mpc$^{-1}$ and 
$q_0 = 0.5$.

\section{The Cluster Local XLF}
The procedure to convert the observed counting rate of the clusters 
to luminosities was the following. 
First, source count rates in the 0.5-2.0 keV band were derived by using the
Steepness Ratio Technique (SRT, Paper I), and converted to the
corresponding un-absorbed total fluxes as described in Paper II.
Next, we computed the cluster rest-frame luminosities for the 
126 (out of 130) sources with available redshifts.
We made $K$-corrections assuming that the typical spectra of clusters in the 
0.5-2.0 keV energy band approximates a power law with energy index 
$\alpha = 0.4$.

\subsection{The method}
The RASS1 Bright Sample has been obtained by performing a cut in SRT
hard-band count rate at 0.25 counts s$^{-1}$ (see Paper II).
Since different regions of the sky have different amounts of Galactic
absorption, the cut in count rate translates into a range of flux limits
varying between 3.05 and 4$\times 10^{-12}$ ergs cm$^{-2}$ s$^{-1}$ over the
sampled area.
The sky coverage, i.e., the amount of sky surveyed at the different flux
limits, is shown in Figure 6 of Paper II. 
We have derived a non-parametric representation of the XLF of clusters based 
on the method described in detail by Avni \& Bahcall (1980) for 
the coherent analysis of a set of independent samples.
This method is a generalization of the classical $1/V_{max}$ estimator 
developed by Schmidt (1968).
The volume $V_a$ that we use corresponds to the volume within which an 
object could have been detected above the flux limits of the sample.
We have divided the observed range of luminosities in a number of bins 
and computed the differential XLF :
$$ {dn\over dL} (L) ~=~ {1\over \Delta L}\sum_{i=1}^{n_j} {1\over V_a(L_i)} 
\eqno(1)$$
where $n_j$ is the number of clusters in the $j^{th}$ bin and $V_a(L_i)$ is 
the available volume corresponding to the luminosity $L_i$. 
The resulting XLF is shown in Figure 1; the $68\%$ error bars are 
determined by using the Marshall (1985) approximation. ~
Errors computed with simple Poissonian statistics are comparable to
those computed with this approximation.

We have also assumed a parametric representation of the XLF and determined
the parameters of the function through a maximum likelihood analysis.
To this end, we have adopted a modified Schechter (1976) expression,
$$ {dn\over dL}(L) ~=~ A L^{-\alpha} \exp (-L/L_*)~ , 
\eqno(2)$$
characterized by the parameters $\alpha$ and $L_*$, which 
determine the shape of the function, and the normalization $A$.
We have fitted the un-binned data to the Schechter function using an extension 
of the maximum likelihood method given in Murdoch, Crawford, \&
Jauncey (1973) and tested 
whether equation (2) is an adequate representation of the data with a 
Kolmogorov-Smirnov (K-S) test (details are given in the Appendix).
To derive $\alpha$ and $L_*$, we have maximized the likelihood function
${\cal L}$ given in the Appendix.
To compute the normalization factor $A$, we have required that the integral 
XLF $n(>L)$ of the Schechter model equal that observed for the minimum 
luminosity of the sample. 

\subsection{The results} 
The maximum likelihood method yielded  best-fit parameters: 
$\alpha = 1.52^{+0.11}_{-0.11}$,
$L_* = 3.80^{+0.70}_{-0.55}\times ~10^{44}$ ergs s$^{-1}$ and
$A = 5.07\pm 0.45\times 10^{-7} $ Mpc$^{-3}$
(10$^{44}$ ergs s$^{-1}$)$^{\alpha -1}$. 
The $68\%$ ($1\sigma$) confidence intervals for $\alpha$ and $L_*$
have been obtained by varying $2{\cal L}$ by 1 with respect to its
maximum value (Maccacaro et al. 1988, and references therein).
Errors for $A$ are derived from the uncertainty in the total number of
clusters. 
A K-S test applied to the luminosity function (see Appendix) confirms 
that the Schechter model is an acceptable representation of the data.
Indeed, we find a probability $P_{K-S}$ of exceeding the D statistic 
under the Null Hypothesis (i.e., ``the data set comes from a distribution 
having the theoretical distribution'') equal to 0.59.
The best fitting Schechter function is plotted in Figure 1 as 
a dashed line.

The XLF data points have not been corrected for the small, i.e.,
$\lesssim 10\%$, incompleteness. 
As discussed in Paper II, the main reason responsible for this 
incompleteness is, most likely, the bias against very extended sources 
(i.e. nearby clusters and groups).
Therefore it should affect mainly the faint-end of the XLF (see also
$\S 2.4$).

\subsection{Test on evolutionary effects for z $\lesssim$ 0.3} 
To check for possible evolutionary effects at the moderate redshifts of 
our sample, we have divided the clusters into two redshift bins.
In order to have subsamples with the same number of objects we have
used as separator the median redshift (i.e., $z=0.0785$) of the total
sample.
In the inset of Figure 1 we report the $68\%$ and $90\%$
($\Delta 2{\cal L} = 2.3$ and $4.61$) confidence regions for the
two parameters $\alpha$ and $L_*$ of the two XLFs derived
with the same maximum likelihood procedure described above.
We find that the best-fit $\alpha$, $L_*$ from the high-redshift subsample 
is enclosed within the $95.4\%$ region of the low-redshift subsample, 
whereas the best-fit $\alpha$, $L_*$ from the low-redshift subsample 
is enclosed within the $99.99\%$ region of the high-redshift subsample.   
From the overlap of the $90\%$ confidence regions we do not find
evidence for evolution in our data.
This is in agreement with what has been found in a similar redshift and 
luminosity range by Ebeling et al. (1997), and in other works on more 
distant cluster samples (e.g., Rosati et al. 1998).

\subsection{Comparison with previous works}
In Figure 2, we compare the RASS1 Bright Sample XLF with independent
determinations of the local XLF compiled both from RASS1 data 
and from deeper X-ray surveys. 

We consider first the BCS XLF (Ebeling et al. 1997), which is also
well fitted in the 0.5-2.0 keV band by a Schechter function (eq. [2]).
Inspection of Figure 2 shows that for $L\gtrsim 2\times 10^{43}$ 
ergs s$^{-1}$ the two XLFs are in good agreement.
The difference between the XLFs below $\sim 2\times 10^{43}$ ergs s$^{-1}$
may reflect the different selection methods of the two samples.
A K-S test between our data and the BCS best-fitting Schechter function 
over our whole luminosity range shows, however, that our data could have 
been drawn from the same population of the BCS sample ($P_{K-S}=0.11$).
The second sample compiled from RASS1 data is the
optically selected sample of poor clusters of galaxies from Burns 
et al. (1996). 
A K-S test performed between our data and their best-fitting power law XLF, 
after rescaling their parameters to $H_0 = 50$ km s$^{-1}$
Mpc$^{-1}$ and taking into account only the luminosity range in common 
with our sample, shows that the model agrees with our data ($P_{K-S}=0.67$).

As mentioned above, in Paper II we have estimated on overall 
incompleteness of our sample of $\sim 10\%$, affecting mainly the 
faint end of the XLF.
A direct comparison of the BCS XLF best-fitting model with ours 
at the luminosity of $6.2\times 10^{42}$ ergs s$^{-1}$, representative
of our lowest luminosity bin, shows a difference of about 13\%. 
This result confirms that the incompleteness at faint luminosities must
be indeed modest. 

In Figure 2 we also compare our XLF with those derived from the
EMSS (Henry et al. 1992) and RDCS (Rosati et al. 1998) samples in their 
nearest redshift shells, i.e., $0.14<z<0.2$ and $0.045<z<0.25$, 
respectively. 
We have converted EMSS luminosities from the $0.3-3.5$ keV to 
$0.5-2.0$ keV energy band assuming a Raymond-Smith thermal spectrum with 
temperature 6 keV and 0.3 solar abundances.~
A two-sided K-S test between the EMSS data and ours, shows that the two 
distributions are not statistically different in the luminosity region 
where they overlap ($P_{K-S}= 0.20$).
A one-sided K-S test between our data and the best-fitting power law of 
the RDCS sample shows that our data is consistent with the RDCS XLF
model ($P_{K-S}= 0.40$).
Both the EMSS and RDCS samples are selected purely by the X-ray 
properties of clusters. 
The good agreement between these independent determinations of the 
local XLF and our result, indicates that our cluster candidates 
pre-selection method, which is also based on optical information 
(Paper II), is well under control.
 
Finally, in Figure 3 we compare our XLF with those from
the EMSS and RDCS, both computed for the $0.3<z<0.6$ redshift range,
plus the RDCS XLF for $0.5<z<0.8$.
Inspection of Figure 3 shows that the RDCS sample, covering a relatively
small solid angle ($\sim 33$ deg$^2$) with respect to the EMSS surveyed 
area (735 deg$^2$), cannot probe the XLF above $2-3\times 10^{44}$
ergs s$^{-1}$, and that, for faint luminosities, no evolution seems to 
be present.
On the contrary, the EMSS sample, being relatively shallower than the 
RDCS, probes mostly the bright-end of the XLF.
We have compared the EMSS XLF in the $0.3<z<0.6$ redshift range with
our XLF in the luminosity range covered by both functions 
($0.74 < L_X < 8.67\times 10^{44}$ ergs s$^{-1}$).
Using a two-sided K-S test between the EMSS data and ours, we find 
a difference of the two which is significant at the $3.3\sigma$ level
($P_{K-S}=8.3\times 10^{-4}$).
This degree of evolution is similar to what was found in Henry et al.
(1992).
Recently, Nichol et al. (1997) have revised the EMSS cluster sample,
using both new X-ray and optical data.
A K-S test between the revised EMSS data and ours in the same luminosity 
range as above gives $P_{K-S}=7.7\times 10^{-4}$, indicating that 
the two XLFs are still different at the $3.3\sigma$ level.
Recently, Reichart et al. (1998) re-examined the EMSS sample and 
noted the deficiency of bright ($L_X > L_*$) high redshift ($z>0.3$) systems.
Also, Vikhlinin et al. (1998a), studying a sample of 
clusters detected in a 160 deg$^2$ survey of $ROSAT$ PSPC fields 
(Vikhlinin et al. 1998b), find evidences of a deficit of high luminosity 
distant clusters at more than $99.5\%$ confidence.
As noted by the authors, their sample was selected by using 
photometric redshifts, and the optical spectroscopic identification 
of the X-ray sources is still in progress.



\acknowledgments
We would like to thank R. Della Ceca, P. Rosati, P. Henry and the referee 
R. Nichol, for their helpful suggestions and discussions.

\appendix
\section{Appendix}
We extend here the maximum likelihood method discussed in Crawford, Jauncey,
\& Murdoch (1970) and Murdoch et al. (1973), which works on the un-binned data 
distribution, in order to estimate the best-fitting parameters of the 
X-ray luminosity function when the assumed model is a Schechter function 
(eq. [2]).
For consistency reasons, we adopt essentially the same notation of 
Murdoch et al. (1973).
We describe the treatment of the multi-volume case for a flux-limited sample
in an error-free situation.
Since the sample discussed in this work is characterized by a set of 
flux-limits, the maximum volume of each source is the {\it available} 
volume $V_a$, as described in Avni \& Bahcall (1980).

In general, if one considers sources within the volume $V_{a_j}$ having
luminosities in the range between $L_{0j}$ and $L_{uj}$, then the probability
density $P(L_{ij})$ for the $i$th source of luminosity $L_{ij}$ in the 
volume $j$ is given by
$$P(L_{ij}) = 
{V_{a_j}\over K} \exp(- L_{ij}/L_*) L_{ij}^{-\alpha},
$$
where the normalization factor $K$ is computed by integrating over all the 
volumes 
$\sum_j \int^{L_{uj}}_{L_{0j}} P(L_{ij}) dL_{ij} \equiv 1$,
which gives
$$K = \sum_j V_{a_j} \int^{L_{uj}}_{L_{0j}} \exp(-L_{ij}/L_*) 
L_{ij}^{-\alpha} dL_{ij}.
$$
The maximum likelihood method consists in maximizing 
the probability density of occurrence 
of the observed values of $L_{ij}$, given the assumed Schechter function.
Following Murdoch et al. (1973), this reduces to maximizing the likelihood
function ${\cal L} = \Sigma_i \Sigma_j ln P(L_{ij})$,
where the summation is over all of the observed luminosities. 
Hence, 
$${\cal L} = \sum_j m_j \ln V_{a_j}  - M\ln K -
 {1\over L_*} \sum_i \sum_j  L_{ij} -
\alpha \sum_i \sum_j \ln L_{ij},
$$
where $m_j$ is the number of sources observed in the volume $V_{a_j}$ and
$M = \Sigma_j m_j$ is the total number of sources in the sample.
The maximum value of ${\cal L}$ is found by constructing a grid of values 
of the likelihood function by varying the parameters $\alpha$ and $L_*$.
The same grid of values is used to compute the uncertainties on the 
two parameters as a function of the confidence level. 

Finally, to test if the Schechter function is an adequate 
representation of the data, we use a K-S test that can 
be applied to unbinned data. 
In this case, we transform the assumed function into a cumulative
distribution ranging from 0 to 1 by computing
$$y_i(L_{ij}^\prime) = \sum_j \int^{L_{ij}^\prime}_{L_{0j}} 
P(L_{ij}) dL_{ij} = 
{\sum_j V_{a_j} \int^{L_{ij}^\prime}_{L_{0j}} 
\exp(-L_{ij}/L_{*}) L_{ij}^{-\alpha} dL_{ij}
\over {\sum_j V_{a_j} \int^{L_{uj}}_{L_{0j}} 
\exp(-L_{ij}/L_{*}) L_{ij}^{-\alpha} dL_{ij}}}
$$
and then examining the observed $y_i$ distribution for departure from the 
expected distribution through the K-S test.
\clearpage
%

\clearpage

%

\clearpage

\begin{figure}
\epsfxsize=\textwidth
\epsffile{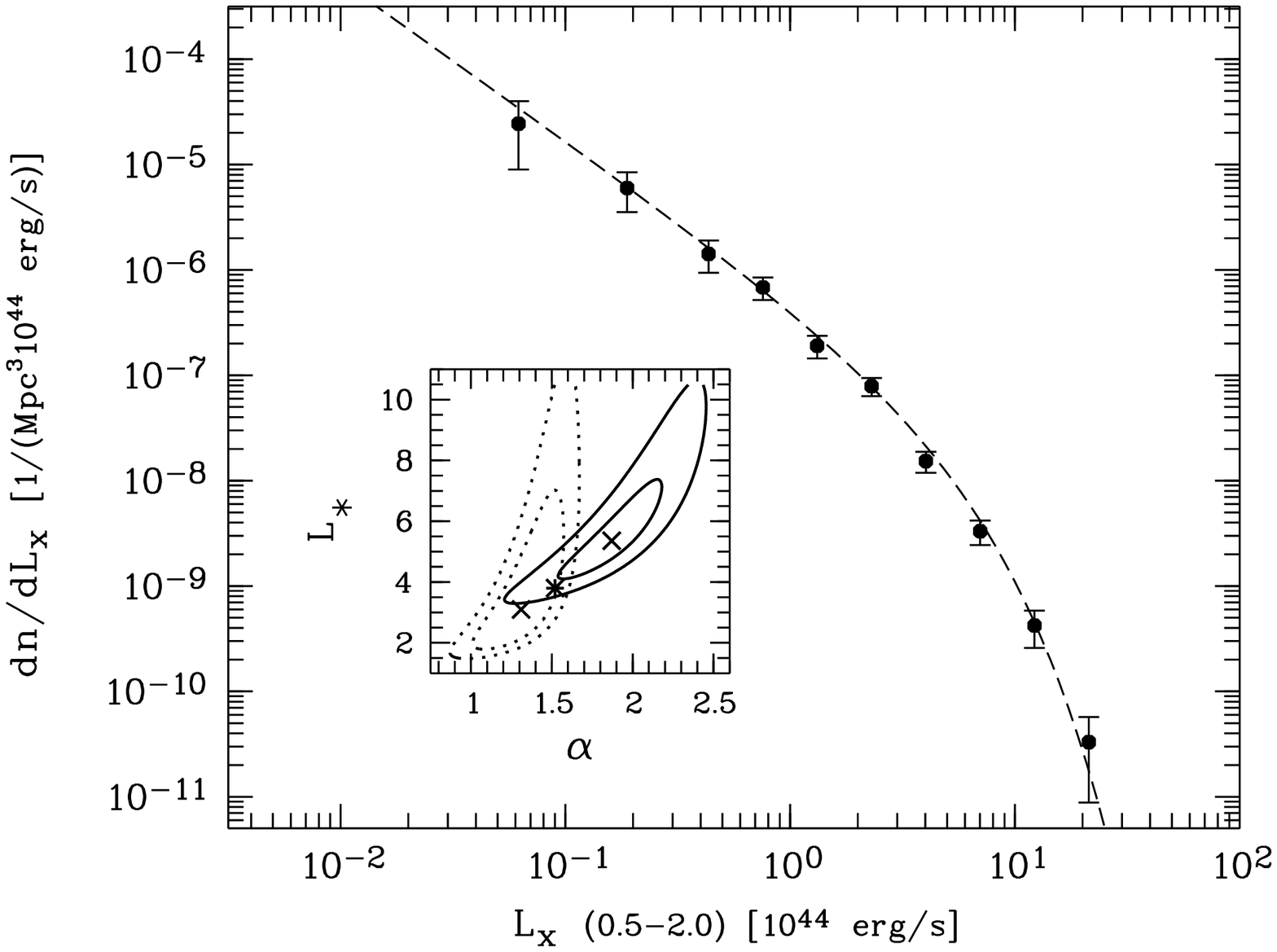}
\vskip -8.truecm
\figcaption {X-ray luminosity function for the RASS1 Bright Sample of
clusters of galaxies. 
Dots and error bars represent the differential XLF, whereas
the long-dashed line is the best-fitting Schechter function estimated 
with a maximum likelihood method from the unbinned data distribution
(see the Appendix). 
The inset shows the $68\%$ and $90\%$ confidence regions for  $\alpha$ and 
$L_*$ (which is in units of $10^{44}$ ergs s$^{-1}$) in two different 
redshift bins: 
dotted and solid lines refer to the $z<z_{median}$ and $z>z_{median}$ 
redshift ranges, respectively.
Crosses refer to the best-fit parameters for the two subsamples, whereas
the asterisk refers to the total sample.
}
\end{figure}

\begin{figure}
\epsfxsize=\textwidth
\epsffile{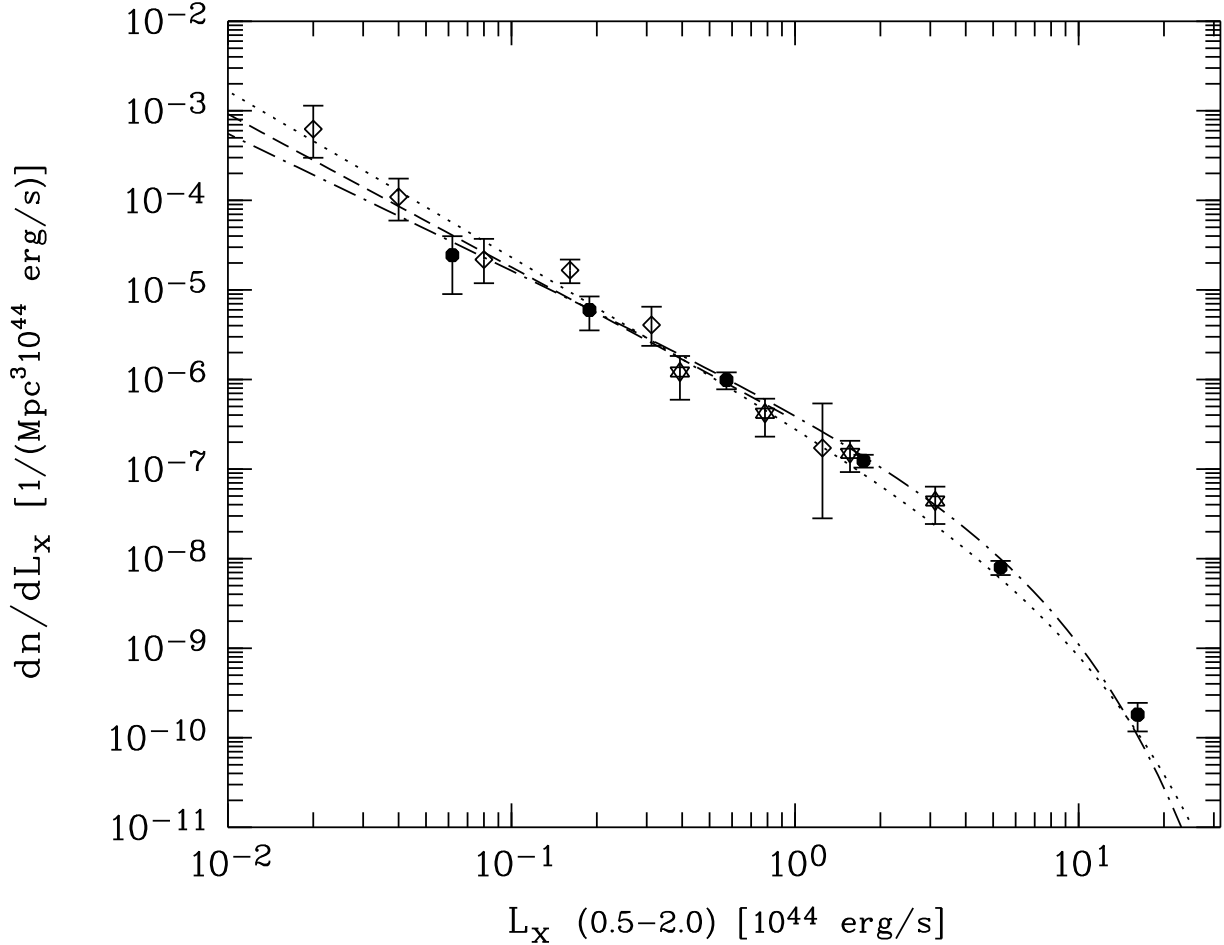}
\vskip -8.truecm
\figcaption {Cluster X-ray luminosity functions in the local universe.
Independent determinations of the local XLF from the $ROSAT$ All-Sky Survey
are shown: filled circles and the dot-dashed line represent the XLF and 
best-fitting Schechter function of the RASS1 Bright Sample (this work), 
the dotted line the BCS XLF (Ebeling et al. 1997), while the long-dashed line 
gives the XLF for poor clusters by Burns et al. (1996).
Also reported are the XLF derived from the local part of deeper surveys:
diamonds for the RDCS (Rosati et al. 1998) and stars for EMSS
(Henry et al. 1992), in the $0.045<z<0.25$ and $0.14<z<0.2$ redshift 
ranges, respectively.
}
\end{figure}

\begin{figure}
\epsfxsize=\textwidth
\epsffile{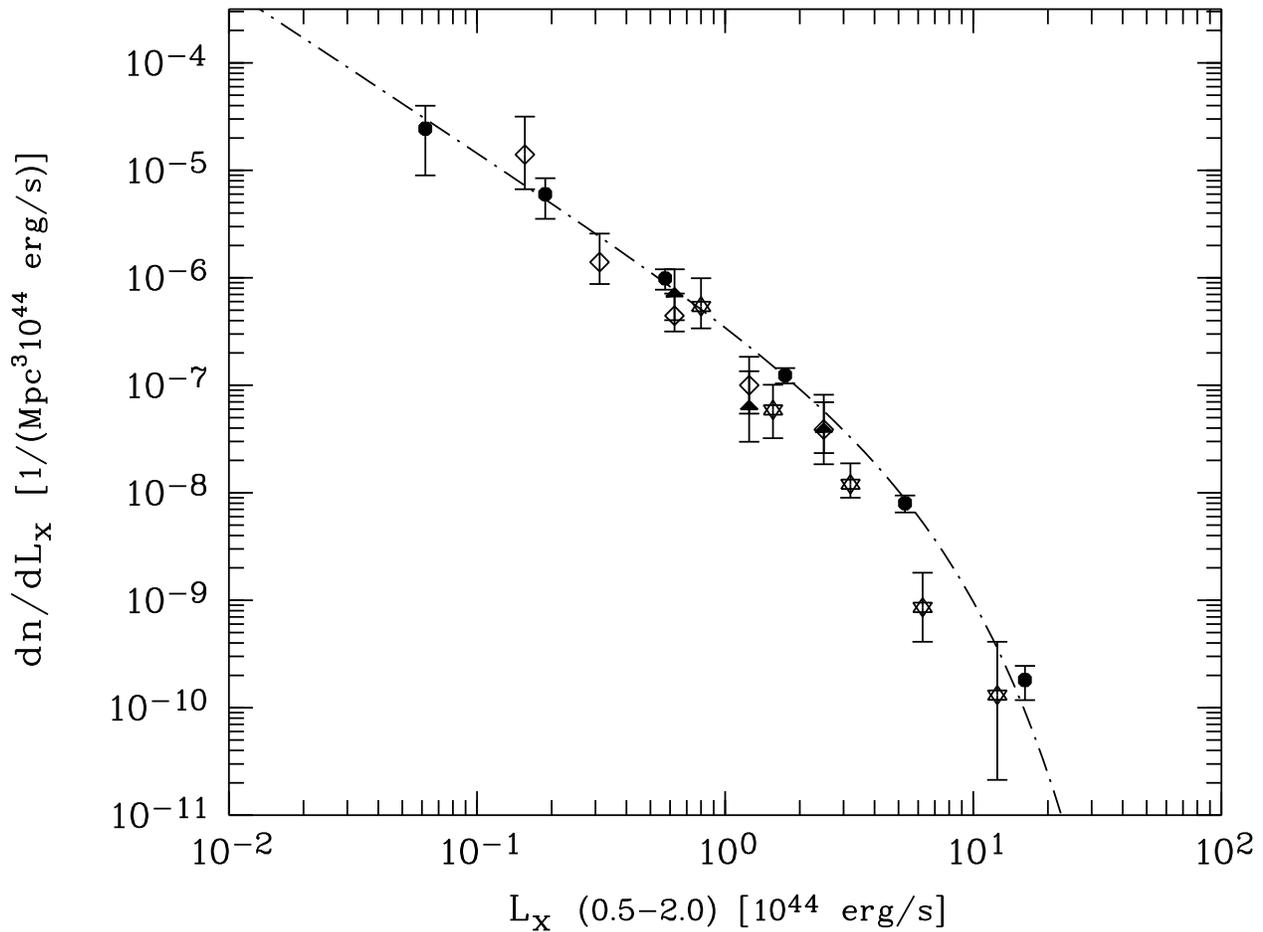}
\vskip -8.truecm
\figcaption {Comparison of the local  and distant cluster X-ray luminosity 
functions. 
Filled circles and the dot-dashed line represent the local XLF and the
best-fitting Schechter function of the RASS1 Bright Sample (this work).
Diamonds and stars are from the RDCS and the EMSS, both computed in the 
$0.3<z<0.6$ redshift range, while filled triangles give the RDCS XLF for 
$0.5<z<0.85$.
}
\end{figure}

\end{document}